\documentclass[twocolumn,showpacs,floatfix,prb,superscriptaddress]{revtex4}
\usepackage{graphicx}
\usepackage{amsmath}
\usepackage{bm}

\begin{document}
\title{Hall conductance from Berry curvature in carbon nanotubes}
\author{J.~D.~Brand}
\affiliation{Department of Physics, Stellenbosch University, Private Bag X1, 7602 Matieland, South Africa}
\author{I. Snyman}
\email{isnyman@sun.ac.za}
\affiliation{Department of Physics, Stellenbosch University, Private Bag X1, 7602 Matieland, South Africa}
\affiliation{National Institute for Theoretical Physics, Private Bag X1, 7602 Matieland, South Africa}
\date{May 2011}
\begin{abstract}
We analytically show that a gap is induced around the Dirac point in the electronic spectrum 
of a previously metallic nanotube, in the presence of electric and magnetic fields perpendicular
to the tube axis. For realistic values of the fields, a gap of at least a few meV can appear.
Despite the quasi-one dimensional nature of the system, the gapped state is associated with
a non-zero topological invariant and supports a Hall effect. This is revealed
when the flux through the tube is varied by one flux quantum, which leads to exactly one electron 
per spin being transported between the ends of the tube.
\end{abstract}
\maketitle

The behavior of electrons in carbon nanotubes under the influence of magnetic fields
has received both theoretical\cite{Aji, Sai04, Lee03, Per07} and experimental\cite{Kan02}
attention. When a strong magnetic field is applied
perpendicular to the tube axis, the Fermi velocity of states around the Dirac points
is reduced. This ``flattening out'' of the dispersion relation
is reminiscent of the dispersionless Landau levels in a 2D sample exposed to a constant magnetic field. 
It is therefore natural to ask whether a carbon nanotube in a perpendicular magnetic field
supports a Hall effect. One type of Hall effect has already been pointed out theoretically\cite{Per07}.
In a strong magnetic field, there are states localized on the tube flanks (i.e. the parts of the tube wall
that are tangential to the magnetic field). States on opposite
flanks carry current in opposite directions. When the Fermi energy lies among these states, an electric field 
perpendicular to both the magnetic field and the tube axis induces a current along the axis of the tube.
 
In this letter we study a different type of Hall effect. In order to observe it,
a gapped state is induced by means of an electric field. 
The mechanism that is responsible for the gap 
is the same as in Ref.\,\onlinecite{Sny09}. When the Fermi energy 
is in this gap, a circumferential electric field induces a current along the tube axis.
At the heart of the effect is the anomalous velocity of Karplus and Luttinger\cite{Kar54} and
the Berry curvature of the band structure of Dirac fermions\cite{Xia10}.    
To make a link with the two dimensional topological invariant that appear in this context, 
a further magnetic field, parallel to the tube axis is introduced. The resulting flux through
the tube plays the role of
crystal momentum in the transverse direction in this one-dimensional system\cite{Niu85}. 

Our main results are the following. Magnetic and electric fields perpendicular
to the tube axis produce a gap around the Dirac point in the electronic spectrum (cf. Eq.\,\ref{mu}). 
For instance, a magnetic field of $30$\,T together with an electric field of $5$\,V/$\mu$m 
produces a gap of $3.7$\,meV in a tube with radius $2.5$\,nm. We derive an expression
for the Hall conductance when the Fermi energy is in the gap (Eq.\,\ref{sigma}). Finally, we show that varying the
flux through the tube by one flux quantum $\phi_0=h/e$, leads to exactly one electron per spin
being transported between the ends of the tube. The argument is the reverse of Laughlin's 
argument for the quantization of the Hall effect\cite{Lau81}. 

These effects occur when electrons are confined to a single 
graphene sheet that is rolled up into a cylinder. 
We would like to vary the flux through the cylinder by
at least one flux quantum. 
In a $20$\,T field parallel to the cylinder axis, this requires a radius
of $8$\,nm or more. 
Single wall tubes found to date have radii $<3$\,nm\cite{Leb02} and are
therefore too narrow to contain a quantum of flux.
Multi-wall carbon nanotubes on the other hand 
have outer radii of several tens of nanometers.
There is experimental evidence\cite{Kan02,Bac99,Bou04} that
at low temperatures, and small bias voltages, 
transport in a multi-wall tube is confined to the outermost wall.
We therefore expect that effects considered in this letter can be observed in multi-wall nanotubes. 

However, since a multi-wall tube is not the cleanest system to display the effect\cite{Bou04},
we also mention another possible experimental method to obtain the system
we consider. The idea is to manufacture an ultra-wide single wall tube starting from 
a multi-wall tube and
manually removing all the inner cylinders. The removal of inner cylinders from 
a $\sim 1$\,$\mu$m section of multi-wall tube has already been achieved experimentally\cite{Cum00}. 
All experiments that we know of, report the
removal of inner cylinders to leave behind a casing of several outer cylinders.
None the less, we see no fundamental obstacles to producing ultra-wide
single wall tubes by this method.  

\begin{figure}[tbh]
\begin{center}
\includegraphics[width=.7\columnwidth]{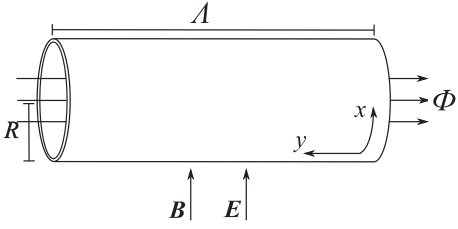}
\caption{The system consists of a single graphene sheet wrapped into a
cylinder of length $\Lambda$ and radius $R$ with $R\ll \Lambda$. A magnetic
field $\bm{B}$ and electric field $\bm{E}$ that are parallel to each other, are
applied perpendicular to the cylinder axis. There is an additional magnetic field
parallel to the cylinder axis, such that the magnetic flux through the
cylinder cross-section is $\Phi$.\label{fsetup}}   
\end{center}
\end{figure}

The system we consider is depicted in Fig.\,(\ref{fsetup}). 
In order to have a tractable problem, we work in the long wavelength
limit, in which it is assumed that states contributing to the Hall conductance
have energies small compared to the hopping amplitude $t\sim2.7\,$eV. 
For the present we consider a single spin species. The effect of including spin
is simply to produce a Zeeman splitting in the single particle spectrum. We
will include this effect eventually.

According to Ref.\,\onlinecite{And05}, electrons in the tube are described by the Dirac equation
\begin{equation} 
\left[v\bm{\sigma}\cdot(-i\hbar\partial_{\bm r}+e\bm{A})+U\right]\Psi=\varepsilon\Psi,\label{dirac1}
\end{equation}
subject to the boundary condition $\Psi(\bm{r}+2\pi R\hat{\bm{x}})=e^{i(\theta-2\pi\nu\tau/3)}\Psi(\bm{r})$
where $R$ is the tube radius.
We have chosen $\hat{\bm{x}}$ to point along the circumference and $\hat{\bm{y}}$
to point along the tube axis. The index $\nu$ is determined by the tube chirality and assumes values 
of $0$ or $\pm1$. (See Ref.\,\onlinecite{And05} for the precise definition.) 
When $\nu=0$ the
tube is metallic, whereas it is semi-conducting if $\nu=\pm1$.
The index $\tau$ distinguishes between valleys, with $\tau=+1$ for valley
$\bm{K}$ and $\tau=-1$ for valley $\bm{K'}$.
The phase $\theta$ in the boundary condition
takes into account 
the effect of the component of the
magnetic field parallel to the tube axis.
It is given by $\theta=2\pi \Phi/\phi_0$, where $\Phi$ is the flux through the tube cross section.
 The vector potential
$\bm{A}(x)=B_0 R \sin(x/R)\hat{\bm{y}}$ corresponds to the component of the
magnetic field perpendicular to the tube wall. 
The scalar potential is
$U=U_0\cos(x/R)$.

We define $\bm{k}=k_x\hat{\bm{x}}+k_y\hat{\bm{y}}$ with
\begin{equation}
k_x=\frac{m}{R}+\frac{\theta-2\pi\nu\tau/3}{2\pi R},\label{q}
\end{equation}   
where $m$ is an integer that labels transverse modes. We make the ansatz that eigenstates are of the form 
$\Psi_{\bm{k}\eta}(\bm{r})=e^{i\bm{k}\cdot\bm{r}}\psi_{\bm{k}\eta}(x)/\sqrt{\Lambda}$ where $\Lambda$ is the
length of the tube. 
 In the physically relevant regime where $\Lambda$
is far larger than all other relevant length scales, results are independent of the boundary condition in
the $y$ direction.
For convenience, we therefore assume
periodic boundary conditions in the $y$ direction, so that $k_y$ is quantized as
$k_y=2\pi n/\Lambda$ with $n$ an integer.
The index $\eta=\pm1$ distinguishes
between positive energy (conduction band) and negative energy (valence band) solutions.
The wave function $\psi_{\bm{k}\eta}$ is normalized to unity, and obeys 
periodic boundary conditions in the circumferential direction. It satisfies 
$H(\bm{k})\psi_{\bm{k}\eta}=\varepsilon\psi_{\bm{k}\eta}$ 
where $H(\bm{k})=H_0+H_1(\bm{k})$ 
and
\begin{subequations}
\begin{align}
&H_0=v\bm{\sigma}\cdot(-i\hbar\hat{\bm{x}}\partial_x+e\bm{A}),\\
&H_1(\bm{k})=\hbar v\bm{\sigma}\cdot\bm{\bm{k}}+U. 
\end{align}
\end{subequations}

To find the eigenstates of $H(\bm{k})$ close to the Dirac point approximately, 
the strategy is the same as in Ref.\,\onlinecite{Sny09}. The two degenerate zero energy eigenstates of $H_0$ are
found exactly. $H_1(\bm{k})$ is then treated as a perturbation. For this to be an accurate approximation,
$\hbar v k_x$, $\hbar v k_y$ and $U_0$ have to be small compared to the level spacing of $H_0$.
In the limit of a weak magnetic field, i.e. $B_0\ll \hbar/eR^2$, the level spacing of $H_0$ is given
by $\hbar v/R$. In this case there is at most one transverse mode $m$, 
namely the one that minimizes $|k_x|$, for which the first order perturbative result is accurate.
(All other modes have $\hbar v |k_x|>\hbar v/R$.) In the opposite limit of a strong magnetic field,
the level spacing of $H_0$ is of the order of the magnetic energy $v\sqrt{e\hbar B_0}$ and the 
perturbation theory is accurate for many transverse modes $m$.    

The two zero energy eigenstates of $H_0$ are
\begin{equation}
f_{+}(x)=N\left(\begin{array}{c}e^{F(x)}\\0\end{array}\right),~
f_{-}(x)=N\left(\begin{array}{c}0\\e^{-F(x)}\end{array}\right),
\end{equation}
where $F=(R/l_m)^2\cos(x/R)$, and $l_m=\sqrt{\hbar/eB_0}$ is
the magnetic length associated with $B_0$. The normalization constant is given by
$N^{-2}=R\,I_0(2R^2/l_m^2)/2 \pi$. To first order in $\bm{k}$ and $U_0$,
the eigenstates of $H(\bm{k})$ are
$\psi_{\bm{k}\eta}(x)=\chi_{\bm{k}\eta}^+f_+(x)+\chi_{\bm{k}\eta}^-f_-(x)$ where the spinors 
$\chi_{\bm{k}\eta}=(\chi_{\bm{k}\eta}^+,\chi_{\bm{k}\eta}^-)^T$ satisfy
\begin{equation}
(\hbar\tilde{v}\bm{\sigma}.\bm{k}+\mu\sigma_z)\chi_{\bm{k}\pm}=\pm\xi\chi_{\bm{k}\eta}.
\end{equation}
The energy of the state is 
\begin{equation}
\varepsilon=\pm\xi=\pm\sqrt{\mu^2+(\hbar\tilde{v}\bm{k})^2}.\label{energy}
\end{equation} 
The mass $\mu$ and renormalized Fermi velocity are respectively given by
\begin{subequations}
\label{muv}
\begin{align}
&\mu=U_0I_1(2R^2/l_m^2)/I_0(2R^2/l_m^2),\label{mu}\\
&\tilde{v}=2\pi RN^2v=v/I_0(2R^2/l_m^2),\label{v}
\end{align}
\end{subequations}
where $I_n(z)$ is a modified Bessel function.
In Appendix \ref{apa}, we test the accuracy and regime of validity of 
Eqs.\,(\ref{energy}) and (\ref{muv}) by comparing to numerics. 

In our analysis, we have neglected the 
Zeeman splitting. When the Zeeman splitting is included, the spectrum is gapped when 
$\mu>\mu_0 B_0$, where $\mu_0=5.788\times10^{-5}$\,eV/T is the Bohr magenton. 
The gap is $\Delta E=2(\mu-\mu_0B_0)$. For small enough $B_0$, $\mu\simeq e U_0 R^2 B_0/\hbar$
is linear in $B_0$. Writing $U_0=eE_0R$ where $E_0$ is the electric field strength,
the condition for a gapped spectrum becomes $E_0>\hbar\mu_0/e^2R^3$.
For a tube with radius $R=2.5$\,nm, a gap appears when $E_0>2.3$ V/$\mu$m. 
For instance, including the Zeeman splitting,
an electric field of $5$\,V/$\mu$m will produce a gap of $3.7$\,meV at $B_0=30$\,T in a tube 
with radius $R=2.5$\,nm. Because the same $E_0$ leads to a larger $U_0$ for larger $R$, the
Zeeman splitting becomes less relevant and at the same time attainable gaps become larger, in tubes
with larger radii. 

The next step is to calculate the Hall conductance when the Fermi energy is in the gap. 
We use the approximate expressions
for the eigenstates and energies we obtained above.
The Kubo formula for the contribution of valley $\tau$ to the Hall conductance reads
\begin{align}
&\sigma_{yx}^\tau=\frac{2e^2}{h}\frac{(\hbar v)^2}{\Lambda R}\int_0^L dx_1\int_0^\Lambda dy_1\int_0^L dx_2\int_0^\Lambda dy_2\,
\sum_{\bm{k}_1,\bm{k}_2}\nonumber\\
&\frac{{\rm Im}\left\{\left[\Psi_{\bm{k}_1-}(\bm{r_1})^\dagger\sigma_y\Psi_{\bm{k}_2+}
(\bm{r_1})\right]\left[\Psi_{\bm{k}_2+}(\bm{r_2})^\dagger\sigma_x\Psi_{\bm{k}_1-}(\bm{r_2})\right]\right\}}{(\varepsilon_1-\varepsilon_2)^2}.
\label{kubo1}
\end{align}
Using the perturbative solutions for $\Psi_{\bm{k}\pm}$,
one finds
\begin{align}  
\int_0^L dx\int_0^\Lambda dy\,\Psi_{\bm{k}_1\eta_1}&(\bm{r})^\dagger\bm{\sigma}\Psi_{\bm{k}_2\eta_2}(\bm{r})\nonumber\\
&=\delta_{\bm{k}_1,\bm{k}_2}2\pi RN^2\chi_{\bm{k}_1\eta_1}^\dagger\bm{\sigma}\chi_{\bm{k}_1\eta_2}.
\end{align}
Substituting the explicit form of $\chi_{\bm{k}\pm}$, the Kubo formula becomes
\begin{equation}
\sigma_{yx}^\tau=\frac{1}{2}\frac{e^2}{h}\frac{\hbar^2\tilde{v}^2}{\Lambda R}\sum_{\bm{k}}\frac{\mu}{\xi^3}.
\end{equation}
The sum $\frac{1}{\Lambda}\sum_{k_y}$ is replaced by the integral $\frac{1}{2\pi}\int_{-\infty}^\infty dk$ to obtain
\begin{equation}
\sigma_{yx}^\tau=\frac{e^2}{h}\frac{\hbar\tilde{v}}{2\pi R}\sum_{k_x}\frac{\mu}{\mu^2+\hbar^2\tilde{v}^2k_x^2}.
\end{equation}
Summing the contributions from the two valleys together and substituting for $k_x$
from Eq.\,(\ref{q}) leads to
\begin{align}
\sigma_{yx}=\frac{e^2}{h}\sum_m\big[&\frac{\gamma}{\gamma^2+(2\pi m+\theta-2\pi\nu/3)^2}\nonumber\\
&\hspace{1cm}+\frac{\gamma}{\gamma^2+(2\pi m+\theta+2\pi\nu/3)^2}\big],\label{sigma}
\end{align}
with $\gamma=2\pi R \mu/\hbar \tilde{v}$. To obtain this result we tacitly made two assumptions.
(1) We assumed that first order perturbation theory in $\bm{k}$ and $U_0$ is sufficiently accurate for 
all modes that contribute to $\sigma_{yx}$. (2) We assumed that modes
included in the sum in Eq.\,(\ref{sigma}) but for which first order perturbation theory is
not accurate, give a negligible contribution. Do these assumption place a restriction on
the magnetic field strengths for which Eq.\,(\ref{sigma}) is valid? 
Comparison with numerical calculations (see below) indicate that the
result is most accurate in the limits of small or large $R/l_m$, and less accurate for intermediate values.

In both the small and large $\gamma$ limits, further simplifications are possible. 
For $\gamma\ll2\pi$, all but the term with the smallest denominator in Eq.\,(\ref{sigma}) 
make negligible contributions to the sum. 
For metallic tubes, $\nu=0$ and the Hall conductance as a function of $\theta$ consists of
a sequence Lorentzian peaks at $\theta=2\pi m$, $m=0,\,\pm1,\,\pm2,\,\ldots$ 
with heights $2e^2/\gamma h$ and widths $\gamma$.
For semi-conducting tubes, $\nu=\pm1$ and there
are two sequences of Lorentzian peaks in the Hall conductance, at $\theta=2\pi m\pm 2\pi/3$, 
$m=0,\,\pm1,\,\pm2,\,\ldots$. The peak heights are $ e^2/\gamma h$ and the peak widths $\gamma$. 

In the opposite limit of $\gamma\gg2\pi$, the sum in Eq.\,(\ref{sigma}) can be converted into an integral, and we
obtain
\begin{equation}
\sigma_{yx}^{\gamma\gg2\pi}=\frac{e^2}{h}.\label{quan1}
\end{equation}
The $\theta$ independence of $\sigma_{yx}$ can be understood as follows.
Eigenstates close to the Dirac point are localized to the scale $l_m$ and therefore
become insensitive to boundary conditions when $l_m<R$. Since $\theta$ only appears
in the boundary condition, the gap is independent of $\theta$.

Eq.\,(\ref{quan1}) shows that the Hall conductance $\sigma_{yx}$ becomes quantized when $R/l_m$
is large enough. Related to this, the average of $\sigma_{yx}(\theta)$ over $\theta$,
i.e. $\left<\sigma_{yx}\right>_\theta=\int_{0}^{2\pi} d\theta\,\sigma_{yx}(\theta)/2\pi$
is always quantized. Using the result of Eq.\,(\ref{sigma}) for $\sigma_{yx}(\theta)$, we obtain
\begin{equation}
\left<\sigma_{yx}\right>_\theta=e^2/h.\label{quantized}
\end{equation}
Although we have only derived this result in the small $U_0$ limit, it holds beyond
this regime. Indeed, it can be shown\cite{Niu85} very generally that $\left<\sigma_{yx}\right>_\theta$
equals an integer multiple of $e^2/h$. This quantization is topologically protected.
    
\begin{figure}[tbh]
\begin{center}
\includegraphics[width=.9\columnwidth]{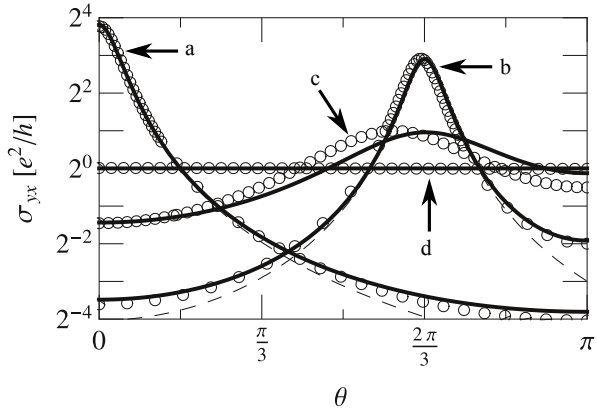}
\caption{The Hall conductance $\sigma_{yx}$ as a function of $\theta$, the magnetic
flux through the tube, in units of $\hbar/e$. The solid curves are 
the analytical result of Eq.\,(\ref{sigma}). The dashed curves for (a) and (b) show 
the single Lorentzian peak approximation, valid for small $\gamma$. 
The circles were calculated numerically using the nearest neighbor tight binding Hamiltonian 
for a zig-zag nanotube with $4000$ hexagons along the 
length of the tube. Thus the tube length is $\Lambda=0.85$\,$\mu$m. 
(a) and (b) show results for weak magnetic fields, i.e. $R/l_m$ small.
For (a) a tube of radius $R=51 a/2\pi$, i.e. with $51$ hexagons around
the circumference, was used. For (b), (c) and (d) a tube with $50$ hexagons around the 
circumference was used. (This corresponds to radii of $1.97$\,nm and $1.93$\,nm respectively.)
In (a) and (b) $U_0=0.01\,t$ and the perpendicular magnetic field strength is $B_0=0.005ea^a/h\simeq53$\,T 
corresponding to a magnetic length $l_m=\simeq3.5$\,nm. 
In (c) and (d) results are shown for intermediate and large $R/l_m$. In (c) and (d), $U_0=0.007\,t$.
In (c) $B_0=0.015\,ea^2/\hbar$ corresponding to a magnetic length $l_m=2.0$\,nm.
In (d) $B_0=0.05\,ea^2/\hbar$ corresponding to a magnetic length $l_m=1.1$\,nm. \label{f2}}   
\end{center}
\end{figure}

Now we compare our analytical results to numerical results obtained using
the nearest neighbor tight binding Hamiltonian of a zig-zag nanotube with $4000$ hexagons
along the length of the tube, giving a tube length $\Lambda=0.85$\,$\mu$m. 
The computation time of the numerical algorithm limits us to considering fairly narrow tubes 
($R\stackrel{<}{\sim}2$\,nm). We therefore have to consider magnetic fields
that are unrealistically large in order to obtain values of $R/l_m\stackrel{>}{\sim}1$. In 
physical realizations $R$ will be larger, allowing the large $R/l_m$ limit to be reached for
experimentally achievable values of $B_0$.
 
In Fig.\,(\ref{f2}) we plot $\sigma_{yx}$ versus $\theta$ for $\theta$ between $0$ and $\pi$. 
(The curves are
symmetric about $\pi=0$ so that we do not have to consider the interval $(-\pi,0)$ separately.)
In (a) and (b) are shown
the analytical results of Eq.\,(\ref{sigma}) (solid lines) and the small $\gamma$ approximation
of a single Lorentzian peak (dashed lines)
compared to numerical results (circles).
In both cases  $U_0=0.01\,t$ and $B_0=0.005\,h/ea^2\simeq53$\,T. 
In (a) the radius was
$R=51\,a/2\pi$ so that $\bm{L}=51\bm{a}$ and $\nu=51_{{\rm mod}\,3}=0$. $R/l_m=0.57$
is small and thus we see a Lorentzian peak at $\theta=0$. In (b) $R=50\,a/2\pi$ so that $\bm{L}=50\bm{a}$ and 
$\nu=50_{{\rm mod}\,3}=-1$. $R/l_m=0.56$ is again small and we see Lorentzian peaks at $\theta=\pm2\pi/3$.
In both (a) and (b) we see good agreement between analytical and numerical results.
In (c) the magnetic field strength $B_0=0.015\,ea^2/\hbar$ is in the intermediate regime ($R/l_m=0.97$).
Here the analytical result does not reproduce the numerical result as well as before.
This is due to the level-spacing of the zero order Hamiltonian being of the same order
as the perturbation for some modes that contribute to the Hall conductance. (d) has $B_0=0.05\,ea^2/\hbar$
which is in the strong magnetic field regime ($R/l_m=1.8$). Eigenstates are localized on the scale of
$l_m$ and therefore insensitive to boundary conditions. The Hall conductance is quantized to $e^2/h$
(per spin) independent of $\theta$. Note that the numerical results in (a) through (d)
all obey the expected quantization of the average $\left<\sigma_{yx}\right>_\theta=e^2/h$
(Eq. \ref{quantized}).   

In order to observe the non-zero Hall conductance and the quantization of its average over $\theta$, 
consider the following experiment. 
Suppose the flux $\Phi$ through the tube is varied. This produces an emf $-d\Phi/dt=(h/e)d\theta/dt$
around the circumference of the tube. Owing to the non-zero Hall conductance, a current 
$I$ is produced along the axis of the tube, such that $I=(h/e)\sigma_{yx}(\theta)d\theta/dt$.
The total charge transported through any cross-section of the tube as the flux is changed
by one flux quantum, is $Q=\int_0^{t_0}dt\,I=(h/e)\int_0^{2\pi}d\theta\,\sigma_{yx}(\theta)$.
From Eq.\,(\ref{quantized}) then follows that $Q=e$, i.e. one electron (per spin) is transported
through the tube as the flux is changed by one flux quantum. This argument is Laughlin's 
argument for the quantization of the Hall effect\cite{Lau81} applied in reverse.

In conclusion, in the presence of parallel magnetic and electric fields that are perpendicular
to the tube axis, a carbon nanotube that was previously metallic develops a gap.
For realistic values of the fields and the tube radius, gaps of at least several meV can be
induced. We derived an expression for the Hall conductance (Eq.\,\ref{sigma}) when 
the Fermi energy is in the gap. The expression agrees well with numerical results from
the nearest neighbor tight binding Hamiltonian. The non-zero Hall conductance leads to quantized
transport. When the flux through the tube is varied by one flux quantum, exactly one electron
per spin is transported between the ends of the tube. 

\appendix
\section{Gap behavior}
\label{apa}

Here we investigate the behavior of the gap as a function of the system parameters.
We also test the accuracy and range of applicability of the perturbative results
by comparing to results obtained by diagonalizing the nearest neighbor tight binding Hamiltonian
for a zig-zag nanotube numerically.

\begin{figure}[tbh]
\begin{center}
\includegraphics[width=\columnwidth]{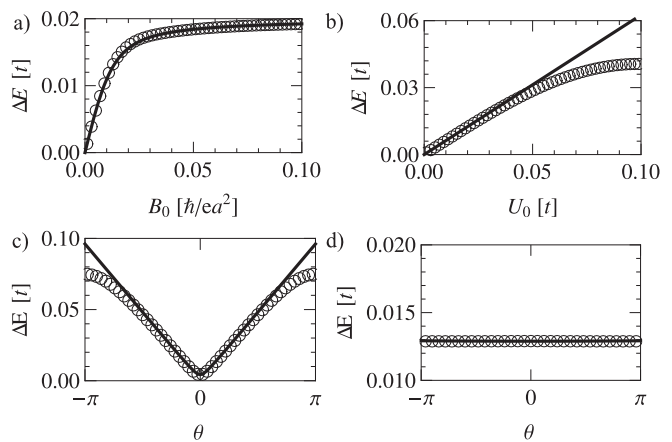}
\caption{Behavior of the gap. All plots are for an infinitely long zig-zag
nanotube with radius $51\,a$. The solid lines are 
obtained from Eqs. (\ref{energy}) and (\ref{mu}). The circles
are obtained by numerically diagonalizing the nearest neighbor
tight-binding Hamiltonian of the tube.
a) The gap $\Delta E$ versus the field strength $B_0$ of the component of the magnetic
field perpendicular to the tube axis. The electrostatic potential strength 
$U_0=0.01\,t$ is small enough for perturbation theory in $U_0$ to be accurate. 
There is no magnetic flux through the tube, so that $\theta=0$.
b) The gap $\Delta E$ versus the scalar potential strength $U_0$. The magnetic field
strength is $B_0=0.005\,\hbar/ea^2$. There is no magnetic flux through the tube, 
so that $\theta=0$.
c) and d) The gap $\Delta E$ versus $\theta$ (the flux though the tube in units of $\hbar/e$).
In both cases $U_0=0.007\,t$. In c) the magnetic field strength is $B_0=0.005\,\hbar/ea^2$
so that the radius $R$ is $0.56$ times than the magnetic length $l_m$
and the gap is still strongly dependent on $\theta$. 
In d) the magnetic field strength is $B_0=0.05\,\hbar/ea^2$
so that the radius is $1.8$ times larger than the magnetic length $l_m$. This is 
sufficiently large for the gap to be independent of $\theta$. \label{f1}}
\end{center}
\end{figure}

Results are plotted in Fig.\,(\ref{f1}).
In (a) the gap $\Delta E$ is plotted as a function of $B_0$.
The solid line shows the perturbative analytical result while the circles show the numerical result.
The value of the scalar potential was taken as $U_0=0.01\,t$. For small $B_0$ the 
small parameter that controls the accuracy of the perturbation expansion is $U_0 R/\hbar v$, which
here has a value of $0.09$. We see that this is sufficiently small for perturbative results
to be accurate. In (b) the gap $\Delta E$ is plotted as a function of the potential $U_0$, 
at $B_0=0.005\,h/ea^2$ and $2\pi R=51\,a$. The analytical result co-incides with the numerical
result for $U_0$ up to $\sim 0.04\,t$, corresponding to a value of $U_0 R/\hbar v=0.37$. 
In (c) and (d) the gap $\Delta E$ is 
plotted as a function of $\theta$. In $(c)$ a value of $B_0=0.005\,\hbar/ea^2$ was used 
corresponding to a ratio $R/l_m=0.56$. Since the magnetic length is longer than the radius,
eigenstates are sensitive to boundary conditions and the gap has a strong $\theta$ dependence.
We also see that when $\theta\sim\pm\pi$, the first order perturbative result is not accurate.
(This is because here the perturbation $\hbar v\bm{\sigma}\cdot\bm{k}$ becomes large 
compared to the level spacing of $H_0$ at small $B_0$.) In (d) a magnetic field $B_0=0.05\,\hbar/ea^2$
was used, which corresponds to a ratio $R/l_m=1.8$.
Eigenstates close to the Dirac point are localized to the scale $l_m$ and therefore
become insensitive to boundary conditions when $l_m<R$. 

\acknowledgments
This research was supported by the National Research Foundation (NRF) of South Africa.


\begin{thebibliography}{99}
\bibitem{Aji} H. Ajiki and T. Ando, J. Phys. Soc. Jpn. {\bf 65}, 505, (1996); ibid. {\bf 62}, 
2470 (1993); ibid. {\bf 62}, 1255, (1993).
\bibitem{Sai04} R. Saito, G. Dresselhaus, and M. S. Dresselhaus, Phys. Rev. B {\bf 50},
14698 (1994).
\bibitem{Lee03} H.-W. Lee and D. S. Novikov, Phys. Rev. B {\bf 68}, 155402, (2003).
\bibitem{Per07} E. Perfetto, J. Gonzalez, F. Guinea, S. Bellucci, and P. Onorato, Phys. Rev. B {\bf 76}, 125430, (2007).

\bibitem{Kan02} A. Kanda, S. Uryu, K. Tsukagoshi, Y. Ootuka, and Y. Aoyagi, Physica B {\bf 323}, 246, (2002).

\bibitem{Sny09} I. Snyman, Phys. Rev. B {\bf 80}, 054303, (2009).

\bibitem{Kar54} R. Karplus and J. M. Luttinger, Phys. Rev. {\bf 95}, 1154, (1954).

\bibitem{Xia10} D. Xiao, M.-C. Chang, and Q. Niu, Rev. Mod. Phys. {\bf 82}, 1959, (2010).

\bibitem{Niu85} Q. Niu, D. J. Thouless, and Y.-S. Wu, Phys. Rev. B {\bf 31}, 3372, (1985).

\bibitem{Lau81} R. B. Laughlin, Phys. Rev. B {\bf 23}, 5632, (1981).

\bibitem{Leb02} S. Lebedkin, P. Schweiss, B. Renker, S. Malik, F. Hennrich, M. Neumaier, C. Stoermer, and M. M. Kappes, Carbon {\bf 40}, 417, (2002).


\bibitem{Bac99} A. Bachtold, C. Strunk, J.-P. Salvetat, J.-M. Bonard, L. Forro, T. Nussbaumer, and C. Sch\"onenberger, Nature {\bf 397}, 673, (1999).

\bibitem{Bou04} B. Bourlon, C. Miko, L. Forr\'o, D. C. Glattli, and A. Bachtold, Phys. Rev. Lett. {\bf 93}, 176806, (2004).

\bibitem{Cum00} J. Cumings and A. Zettl, Science {\bf 289}, 602, (2000); Phys. Rev. Lett. {\bf 93}, 086801, (2004).

\bibitem{And05} T. Ando, J. Phys. Soc. Jpn. {\bf 74}, 779, (2005).


\end{thebibliography}
\end{document}